\documentclass[11pt]{article}

\usepackage{amsmath,amssymb}
\usepackage{slashed}
\usepackage{xcolor} 
\usepackage{graphicx}
\usepackage{url}
\usepackage{multirow}
\usepackage[noadjust]{cite}
\usepackage{filecontents}
\usepackage{lineno}
\newcommand{\be}{\begin{equation}}
\newcommand{\ee}{\end{equation}}
\newcommand{\bea}{\begin{eqnarray}}
\newcommand{\eea}{\end{eqnarray}}
\def\Re{{\cal R \mskip-4mu \lower.1ex \hbox{\it e}\,}}
\def\Im{{\cal I \mskip-5mu \lower.1ex \hbox{\it m}\,}}

\def\tev{\,{\ifmmode\mathrm {TeV}\else TeV\fi}}
\def\gev{\,{\ifmmode\mathrm {GeV}\else GeV\fi}}
\def\mev{\,{\ifmmode\mathrm {MeV}\else MeV\fi}}
\def\to{\rightarrow}

\begin{document}

\renewcommand{\thefootnote}{\fnsymbol{footnote}}
\begin{center}
\vspace*{15mm}

\vspace{0.2cm}
{\Large \bf  Search for top quark flavour changing neutral currents in same-sign top quark production} \\
\vspace{1cm}

{\bf  Reza Goldouzian\footnote{email: r.goldouzian@ipm.ir} }

 \vspace*{.1cm}
   {\small\sl School of Particles and Accelerators, Institute for Research in Fundamental Sciences (IPM) \\ P.O. Box 19395-5531, Tehran, Iran }

\end{center}

\vspace*{2mm}
\begin{abstract}
The presence of the anomalous top-quark flavour-changing neutral-current (FCNC)  interactions leads to the production of same-sign top quarks in proton-proton collisions. The results of a search for events with same-sign dileptons and bjets conducted by CMS Collaboration with 10.5 fb$^{-1}$ of data collected in $pp$ colisions at  $\sqrt{s}=8$ TeV are used to obtain the constraints on the strength of top-quark FCNC interactions. The 95$\%$ confidence level upper limits on the branching ratios of top-quark decays to a light quark $q=u,c$ and a gauge or a Higgs boson are set to be BR$(t\rightarrow u \gamma) < 1.27\%$, BR$(t\rightarrow u Z) < 0.8\% $, BR$(t\rightarrow u g) < 1.02\% $ and BR$(t\rightarrow u H) < 4.21\% $. The sensitivity of future searches in the same-sign top-quark channel is also presented.

\end{abstract}

\section{Introduction}
\label{sec:1}

Because of the large mass of the top quark near to the electroweak symmetry-breaking scale, the study of top-quark properties can open a unique window to new physics \cite{newphysics}. In the Standard Model (SM) framework, the flavour-changing neutral-current (FCNC) processes are forbidden at tree level and are suppressed at the level of quantum loop corrections due to the Glashow-Iliopoulos-Maiani (GIM) mechanism \cite{gim}. Whereas the SM predicts tiny branching ratios of top-quark FCNC decays to a light up-type quark and a gauge or Higgs boson \cite{smbr} [BR$(t\rightarrow qX) \sim 10^{-17} - 10^{-12}$, where $q$ = up or charm quark and $ X$ = photon ($\gamma$) , $Z$ boson ($Z$), gluon ($g$), or Higgs boson ($H$)],  various extensions of the SM predict a huge enhancement for the branching ratios of these decays by relaxing the GIM suppression and introducing new particles that contribute in the quantum loops \cite{Cao:2014udj,Dedes:2014asa,susybr,susybr2,2hdmbr,technicolorbr}. It indicates that the observation of any sign from top-quark FCNC processes will give evidence of physics beyond the SM. 

\begin{table}[Hb]
\centering
\begin{tabular}{|l|l|l|l|l|}
\hline
 & CDF  &  D0   &  ATLAS   &  CMS       \\ \cline{1-1}
\hline
${\cal B}R(t\rightarrow q\gamma)\% $ & $3.2$   \cite{cdf1} &  - &  - &  $ 0.016$\cite{cmsgamma}  \\
\hline
${\cal B}R(t\rightarrow qZ)\% $ & $3.7$   \cite{cdf2} & $3.2 $\cite{d01} &  $0.73 $\cite{atlasz}  &  $0.05$\cite{cmsz}  \\
\hline
${\cal B}R(t\rightarrow qg)\%  $ & $0.039$\cite{cdfg} &  $0.02$\cite{d0g} &  $0.0031$\cite{atlasg} &   $0.035 $\cite{cmsg} \\
\hline
${\cal B}R(t\rightarrow qH) \% $ & - &  -  &$0.79$\cite{atlash} &  $0.56$\cite{cmsh} \\
\hline
\end{tabular}
\caption{\label{tab:i} The most stringent experimental bounds on FCNC branching ratios obtained in Tevatron and LHC experiments.
}
\label{result}
\end{table}

Over the years, different experiments have searched for FCNC processes in the anomalous decays of top quarks in $t \bar{t}$ events or anomalous productions of single top events. They have observed no clear evidence of the presence of the FCNC processes and the exclusion limits are set on the branching ratios of anomalous top decays. In Table \ref{result} the most stringent limits obtained in hadron colliders from different sensitive channels are shown. Although the SM predicts top-quark anomalous branching ratios many orders of magnitude below the current experimental limits, experiments are closing in on the regions that are available to physics beyond the SM.

\begin{figure}[tbp]
\centering 
\includegraphics[width=0.45\textwidth]{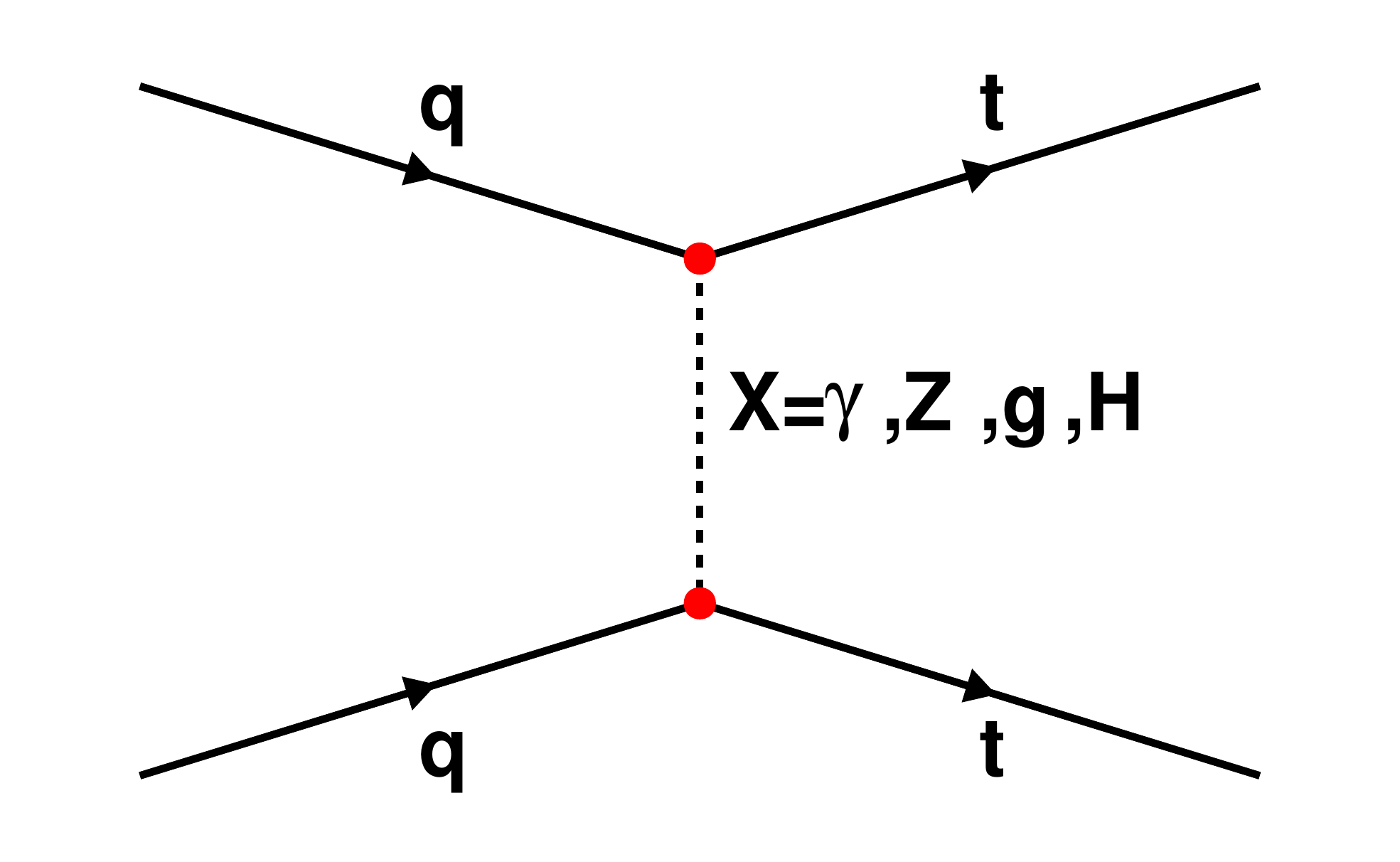}
 \\
\includegraphics[width=0.23\textwidth]{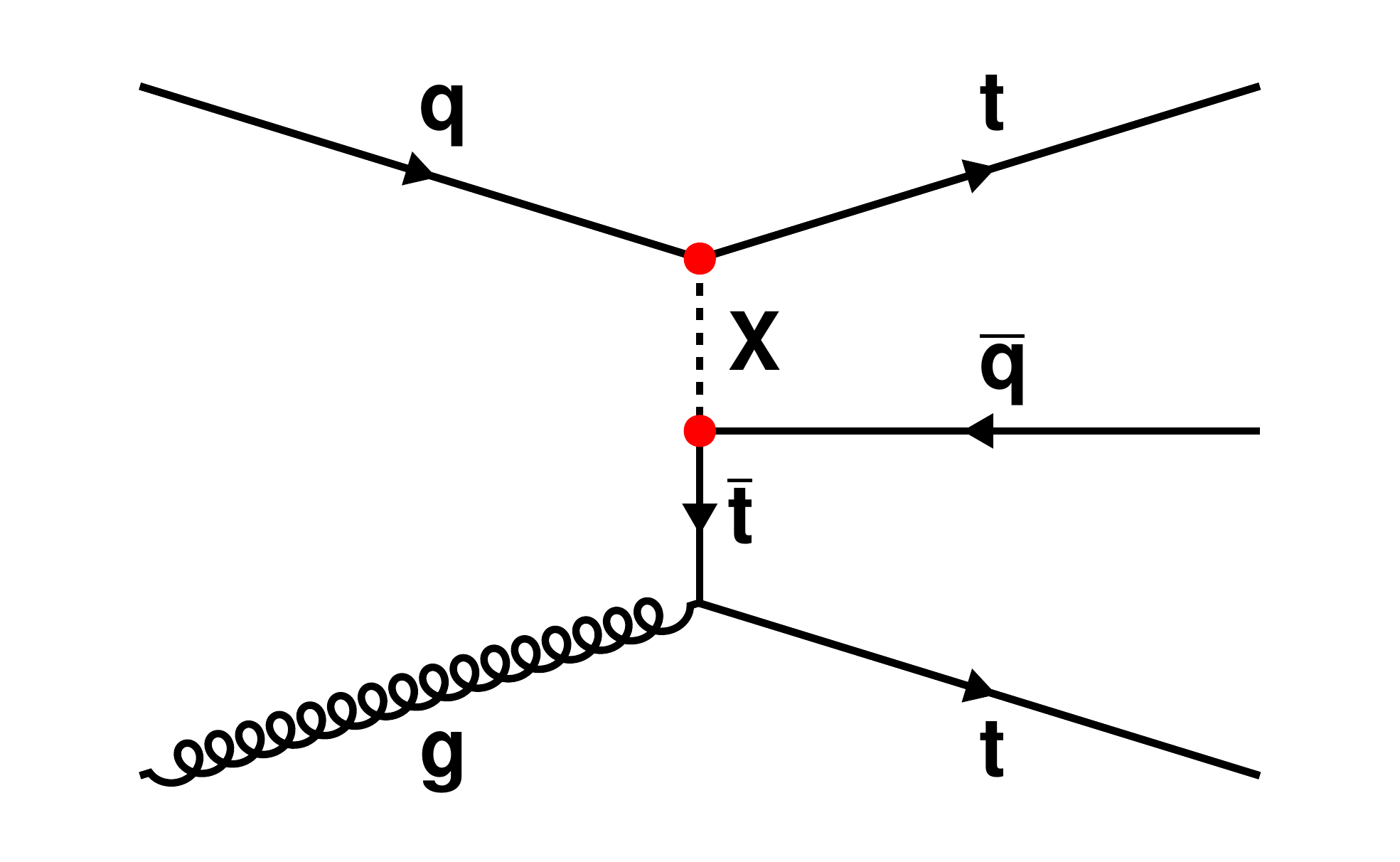}
\includegraphics[width=0.23\textwidth]{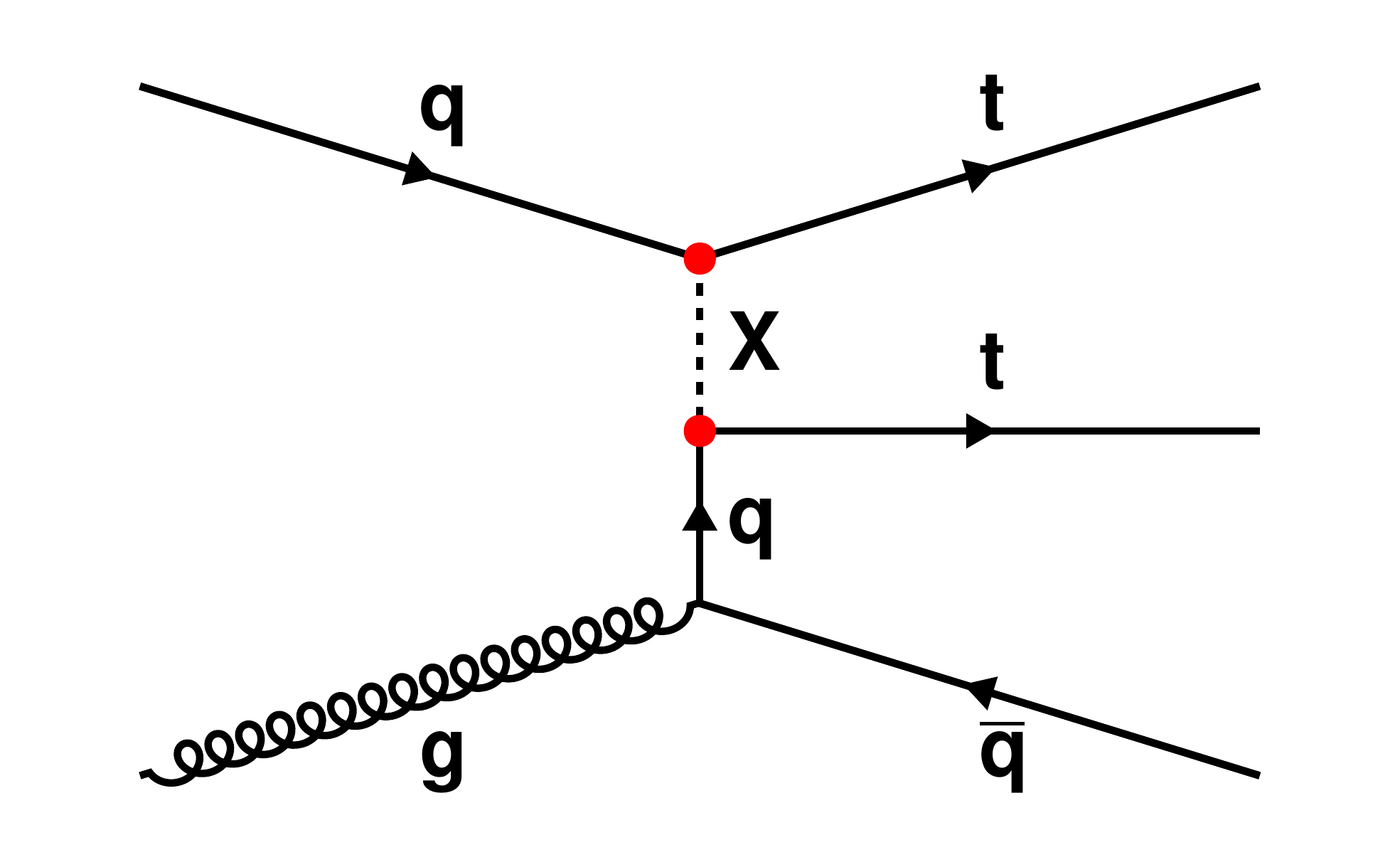}
\includegraphics[width=0.23\textwidth]{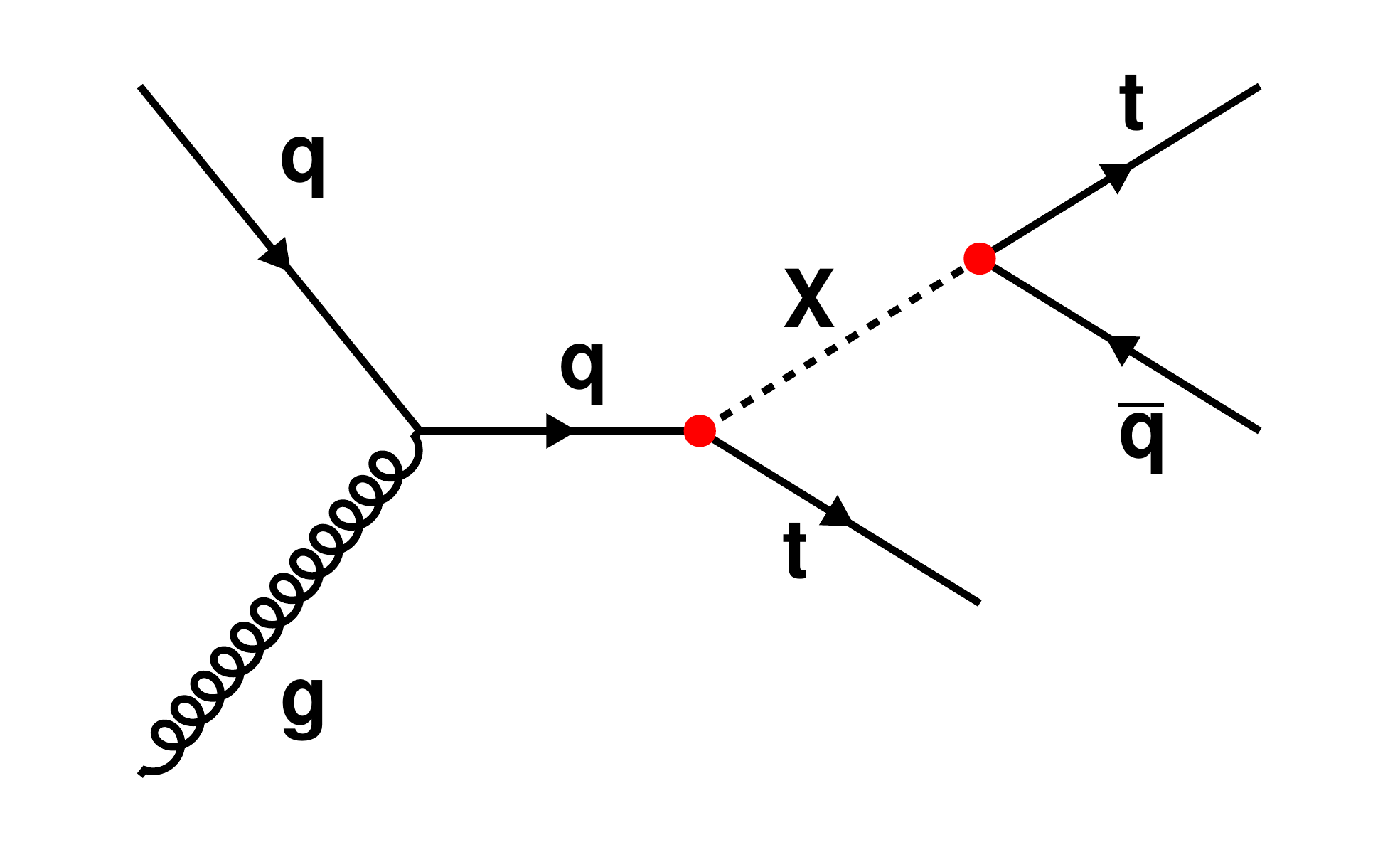}
\includegraphics[width=0.23\textwidth]{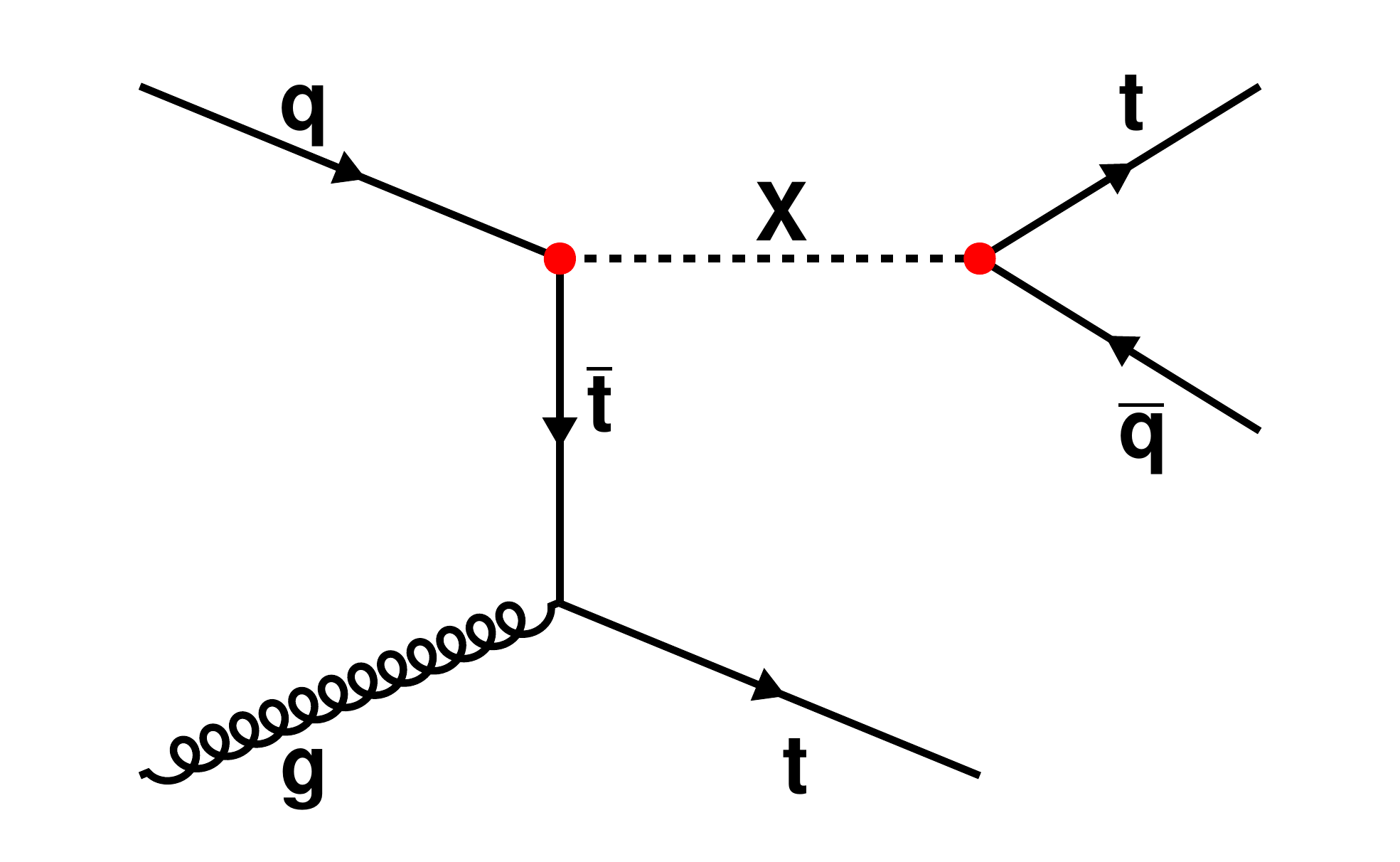}
\hfill
\caption{\label{fig:i} Feynman diagrams describing the production of same-sign top-quark productions (top) and same-sign top + $\bar{q}$ production (bottom) representative of  same-sign top + 1 jet diagrams due to FCNC interactions ($q=u$ or $c$).}
\label{feyn}
\end{figure}

In addition to the anomalous production or decays of top quarks, the FCNC interactions can result in the appearance of a same-sign top quark in hadron colliders \cite{doubletop,doubletophiggs,KhatiriYanehsari}. Figure \ref{feyn} displays the representative diagrams describing the anomalous same-sign top-quark production.  
Same-sign top production followed by the leptonic decay of a $W$ boson from top decays gives rise to final state with same-sign leptons and $b$ jets. Despite the small cross section of the signal channels due to the presence of two anomalous vertices for t$t$ production, this final state has been shown to have very little SM background and is sensitive to new-physics effects \cite{cmstt,FBtt}. Therefore, a same-sign dilepton final state would provide a new window for searching for FCNC interactions.

To investigate the utility of same-sign top production in searching for FCNC interactions, we make use of the results of a search for new physics in events with same-sign dileptons and $b$ jets performed with 10.5 fb$^{-1}$ of data collected from 8 TeV $pp$ collisions by the CMS Collaboration to estimate the upper limit on the strength of the top FCNC anomalous couplings \cite{samesignDilepton}.

 In this work, we study various processes that contribute to the same-sign top-quark final state through $tqX$ vertices, where $X = H$, $\gamma$, $Z$ or gluon. We limit the strength of the FCNC anomalous couplings by considering the leptonic decay of a $W$ boson from top-quark decay and using same-sign dilepton experimental results.
 
The organization of this paper is as follows. Section \ref{sec:2} describes the theoretical framework used to search for FCNC processes. In Sec. \ref{sec:3} we review the CMS same-sign dilepton search and the simulation details of signal samples. The results of the same-sign dilepton search are interpreted in terms of the strength of FCNC interactions in Sec. \ref{sec:4}. The prediction of $95\%$ C.L. exclusion limits at the  14 TeV LHC in Sec. \ref{sec:5} is followed by a conclusion in Sec. \ref{sec:6}.

\section{ Anomalous flavour-changing top-quark couplings }
\label{sec:2}
Top-quark anomalous interactions can be described in a model-independent way by an effective Lagrangian \cite{smbr}. The most general effective Lagrangain describing the interactions between the top quark and a light up-type quark ($u$ or $c$) and a gauge or Higgs boson emerging from dimension-six operators can be written as
\begin{eqnarray}
\label{lagrangy}
-\mathcal{L}_{eff} &=& e\kappa_{q\gamma} \bar{q}\frac{i\sigma^{\mu\nu}q_{\nu}}{\Lambda} [\gamma_{L} P_{L}+\gamma_{R} P_{R}]tA_{\mu} 
\nonumber\\
&+& \frac{g}{2cos{\theta_{w}}}\kappa_{qZ} \bar{q}\frac{i\sigma^{\mu\nu}q_{\nu}}{\Lambda} [z_{L} P_{L}+z_{R} P_{R}]tZ_{\mu} \nonumber\\
&+& g_{s}\kappa_{qg}\bar{q}\frac{i\sigma^{\mu\nu}q_{\nu}}{\Lambda} [g_{L} P_{L}+g_{R} P_{R}]T^a tG_{a\mu}  \nonumber\\
&+& \kappa_{qH}\bar{q} [h_{L} P_{L}+h_{R} P_{R}]tH + H.c.,
\end{eqnarray}

where $e$ is the electron charge, $g$ is the weak-coupling constant, $g_{s}$ is the strong-coupling constant, $\theta_{w}$ is the Weinberg angle, $P_{L,R}=\frac{1}{2} (1\mp \gamma^5)$, $\sigma^{\mu\nu} = \frac{1}{2}[\gamma^{\mu},\gamma^{\nu}]$, and the symbols $\bar{q}$ and $t$ represent the up- (or charm-) and top-quark spinor fields. The parameters $\kappa_{q\gamma}$, $\kappa_{qZ}$, $\kappa_{qg}$ and $\kappa_{qH}$ define the strength of the real and positive anomalous couplings for the current with a photon, $Z$ boson, gluon. or Higgs boson, respectively. The relative contribution of the left and right currents are determined by $\gamma_{L,R}$, $z_{L,R}$, $g_{L,R}$, and $h_{L,R}$, which are normalized as $|\gamma_{L}|^2 + |\gamma_{R}|^2=1$, $|z_{L}|^2 + |z_{R}|^2=1$, etc. 
In the Lagrangian, $q$ is the momentum of the gauge or Higgs boson and
$\Lambda$ is the new physics cutoff which by convention is set to the top-quark mass.

In the literature, there are many alternatives for normalizing the coupling constants in $\mathcal{L}_{\text{eff}}$. Therefore, we will use top-quark branching ratio to express our results to make it comparable with other experimental results. The tree-level prediction for the top-quark decay rate to the $W$ boson and massless $b$ quark is \cite{smbr}
\begin{equation}
\Gamma (t \rightarrow Wb) = \frac{\alpha}{16 s_w^2} |V_{tb}|^2 \frac{m_{t}^3}{m_{W}^2} \left[1-3\frac{m_{W}^4}{m_{t}^4} +2\frac{m_{W}^6}{m_{t}^6}\right]
\end{equation}
The partial decay widths of the top quark with flavor-violating interactions are given by
\begin{eqnarray}
&&\Gamma (t \rightarrow q\gamma) = \frac{\alpha}{4} m_t^3 \frac{|\kappa_{q\gamma}|^2}{\Lambda^2} \nonumber\\
&&\Gamma (t \rightarrow qZ) = \frac{\alpha}{32 s_W^2 c_W^2} m_t^3 \frac{|\kappa_{qZ}|^2}{\Lambda^2}  
\left[1-\frac{m_{Z}^2}{m_{t}^2}\right]^2 \left[2+\frac{m_{Z}^2}{m_{t}^2}\right] 
\nonumber\\
&&\Gamma (t \rightarrow qg) = \frac{\alpha_s}{3} m_t^3 \frac{|\kappa_{qg}|^2}{\Lambda^2} \nonumber\\
&&\Gamma (t \rightarrow qH) = \frac{1}{32 \pi} m_t |\kappa_{qH}|^2 \left[1-\frac{m_{H}^2}{m_{t}^2}\right]^2
\end{eqnarray}

For numerical calculations we set $m_{t}=172.5$ GeV ,$m_{Z}=91.2$ GeV ,$m_{H}=125$ GeV , $s_W^2=0.234$, $\alpha_s= 0.108$, and $\alpha= 1/128.92$.
 
\section{A same-sign top production search for top-quark FCNC interactions}
\label{sec:3}

\subsection{Experimental input}
\label{sec:31}

\begin{table*}[htbH]
\begin{center}
\scalebox{0.7}{
\begin{tabular}{|l|c|c|c|c|c|c|c|c|c|c|c}
\hline
 & SR0 & SR1 &  SR2  &  SR3 & SR4 & SR5 &  SR6  &  SR7 & SR8\\
\hline 
No. of jets ($b$ jets) &$\geq2 (\geq2)$ & $\geq2 (\geq2)$& $\geq2 (\geq2)$ &  $\geq4 (\geq2)$ &$\geq4 (\geq2)$&  $\geq4 (\geq2)$ &  $\geq4 (\geq2)$ & $\geq3 (\geq3)$&$\geq4 (\geq2)$\\
lepton charges &$++/--$ & $++/--$&  $++$ &  $++/--$ & $++/--$&  $++/--$ &  $++/--$ & $++/--$& $++/--$\\
$E_T^{\text{miss}}$ &$\geq0$ GeV & $\geq30$ GeV& $\geq30$ GeV&  $\geq120$ GeV & $\geq50$ GeV& $\geq50$ GeV & $\geq120$ GeV &$\geq50$ GeV& $\geq0$ GeV\\
$H_T$ &$\geq80$ GeV & $\geq80$ GeV&  $\geq80$ GeV & $\geq200$ GeV & $\geq200$ GeV& $\geq320$ GeV & $\geq320$ GeV& $\geq200$ GeV&$\geq320$ GeV\\
\hline
No. of BG &$40\pm14$ & $32\pm11$& $17.7\pm6.1$ &  $2.2\pm1.0$ &$8.1\pm3.4$&  $5.7\pm2.4$ & $1.7\pm0.7$ & $1.2\pm0.6$& $8.1\pm3.3$\\
No. of data &$43$&  $38$& $14$ &$1$ & $10$& $7$ &  $1$& $1$& $9$\\
No. of NP($30\%$ unc.)  &$30.4$ & $29.6$&  $10.7$ & $3.8$ & $12$& $9.6$ & $3.9$ & $4$& $10.5$\\
\hline
\hline
\end{tabular}}
\end{center}
\caption{\label{tab:i} A summary of the results. For each signal region (SR) the kinematic requirements, the prediction for the total background (BG), and the observed number of events are shown. The 95\% C.L.
upper limit on the number of new-physics events under the assumptions of 30\%
uncertainty on the signal efficiency is shown in the last row. Note that the number of jets in the first line of the table includes both b-tagged and non $b$-tagged jets.}
\label{expinput}
\end{table*}

Same-sign dilepton searches at hadron colliders can provide great sensitivity for probing many new-physics models \cite{NP1,FBcorrFCNC,NP2,NP3}. In this work we follow and use the results of the same-sign dilepton and $b$-jets search strategies adopted by CMS Collaboration \cite{samesignDilepton}.

In the analysis two isolated same-sign leptons ($e$ or $\mu$) with $P_{T}>20$ and $|\eta| <2.4$ ($1.442<|\eta|<1.566$ is excluded for the electron) are required. More criteria on the events with a third lepton are applied to minimize the contribution of backgrounds with $\gamma^* \rightarrow l^+ l^-$ and low-mass bound-state and multiboson production. In the CMS report, the lepton identification efficiency, isolation cuts, and detector effects were combined. The lepton selection efficiency is parametrized as \cite{samesignDilepton}
 \begin{equation}
\label{eq:y:3}
\epsilon = \epsilon_\infty \text{erf}  \left(\frac{p_T - 20 GeV}{\sigma}\right) + \epsilon_{20} \left[1-\text{erf} \left(\frac{p_T - 20 GeV}{\sigma}\right)\right]   
\end{equation}
with $\epsilon_\infty=0.65 \ (0.69)$, $\epsilon_{20}=0.35 \ (0.48)$, and $\sigma=42$ GeV ($25$ GeV) for electrons (muons).

Jets are clustered using the anti-$k_t$ algorithm \cite{kt}. At least two jets with $P_{T}>40$ and $|\eta| <2.4$ are needed. $b$ tagging is defined using the combined secondary vertex which uses the information of the secondary vertex and track-based lifetime \cite{csv}. The $b$-tag efficiency is evaluated to be 0.71 for the $b$ jets with $90<p_T<170$ GeV and at higher (lower) $p_T$ it decreases linearly with a slope of $-0.0004 \ (-0.0047) \text{GeV}^{-1}$ \cite{samesignDilepton}. Candidate events are required to have at least two $b$-tagged jets. Finally, in different signal regions different cuts are applied on the scalar sum of the transverse momenta of jets ($H_T$) and the missing transverse energy ($E_T^{\text{miss}}$).

This search divides same-sign dilepton events into several categories, based on the charge of the leptons, the number of selected $b$ jets, and number of selected jets, $H_T$ and $E_T^{\text{miss}}$. Table \ref{expinput}  shows the kinematic requirements, the total background, the observed data, and the upper limit on the number of new-physic events of nine signal regions. The signal regions are not independent and have some overlap with one another, so one cannot combine the limits from different regions.

\subsection{Signal channels and simulation details}
\label{sec:32}

The presence of FCNC interactions leads to the production of $tt$ and $\bar{t} \bar{t}$ through $tuX$ or $tcX$ interactions in proton-proton collisions.  The Lagrangian in Eq. (\ref{lagrangy}) is  implemented in {\sc FeynRules} \cite{feynrule} and
passed to  {\sc MadGraph 5} \cite{madgraph} framework by means of the UFO model \cite{ufo}. The implementation of the leading-order cross section calculated by {\sc MadGraph} is validated for various couplings by comparing the $tX$ production cross sections calculated by {\sc protos} \cite{protos}. 

Due to the larger parton distribution function (PDF) of the $u$ quark (valance quark) in the proton compared to the $c$, $\bar{c}$, and $\bar{u}$ quarks (sea quarks), the main contribution to the cross section comes from the $tt$ production through an anomalous $tuX$ interaction, and the contributions of other signal channels are small \cite{Khatibi}. In Table \ref{crosssection} the cross sections of different signal channels are compared, while CTEQ6L1 is used to evaluate the parton densities.

\begin{table*}[ht]
\centering
\begin{tabular}{|l|l|l|}
\hline
 & $p p \rightarrow t t$ & $p p \rightarrow \bar{t} \bar{t}$ \\ \hline
$tq\gamma$ & $101.25\kappa_{u\gamma}^4$ + $9.60\kappa_{u\gamma}^2$ $\kappa_{c\gamma}^2$ + $0.12\kappa_{c\gamma}^4$ (pb)&  $0.8\kappa_{u\gamma}^4$ + $0.65\kappa_{u\gamma}^2$ $\kappa_{c\gamma}^2$ + $0.12\kappa_{c\gamma}^4$ (pb)\\ \hline
$tqZ$ & $179.85\kappa_{uZ}^4$ + $15.17\kappa_{uZ}^2$ $\kappa_{uZ}^2$ + $0.22\kappa_{cZ}^4$  & $1.3\kappa_{uZ}^4$ + $1.01\kappa_{uZ}^2$ $\kappa_{uZ}^2$ + $0.22\kappa_{cZ}^4$ (pb)\\ \hline
$tqg$ & $44.35\kappa_{ug}^4$ +$4.57\kappa_{ug}^2$ $\kappa_{cg}^2$ + $0.05\kappa_{cg}^4$ (nb) & $0.4\kappa_{ug}^4$ +$0.32\kappa_{ug}^2$ $\kappa_{cg}^2$ + $0.05\kappa_{cg}^4$ (nb) \\ \hline
$tqH$ & $14.02\kappa_{uH}^4$ + $ 4.45\kappa_{uH}^2$ $\kappa_{cH}^2$ + $0.11\kappa_{cH}^4$ (pb)&  $0.52\kappa_{uH}^4$ + $0.47\kappa_{uH}^2$ $\kappa_{cH}^2$ + $0.11\kappa_{cH}^4$ (pb)\\ \hline
\end{tabular}
\caption{\label{tab:i} Same-sign top production cross section due to an anomalous photon, $Z$ boson, gluon, or Higgs exchange at the LHC for $\sqrt{s}=8$ TeV as a function of anomalous couplings. No cut is implemented on the final-state top quarks. CTEQ6L1 is used to evaluate the parton densities, while the renormalization scale $\mu_{R}$ and factorization scale $\mu_{F}$ are fixed at $\mu_{R} = \sqrt{\hat{s}}= \mu_{F}$. Note that the ($p p \rightarrow t t $+ jet) processes are not included. }
\label{crosssection}
\end{table*}

\begin{figure}[tbp]
\centering 
\includegraphics[width=.6\textwidth]{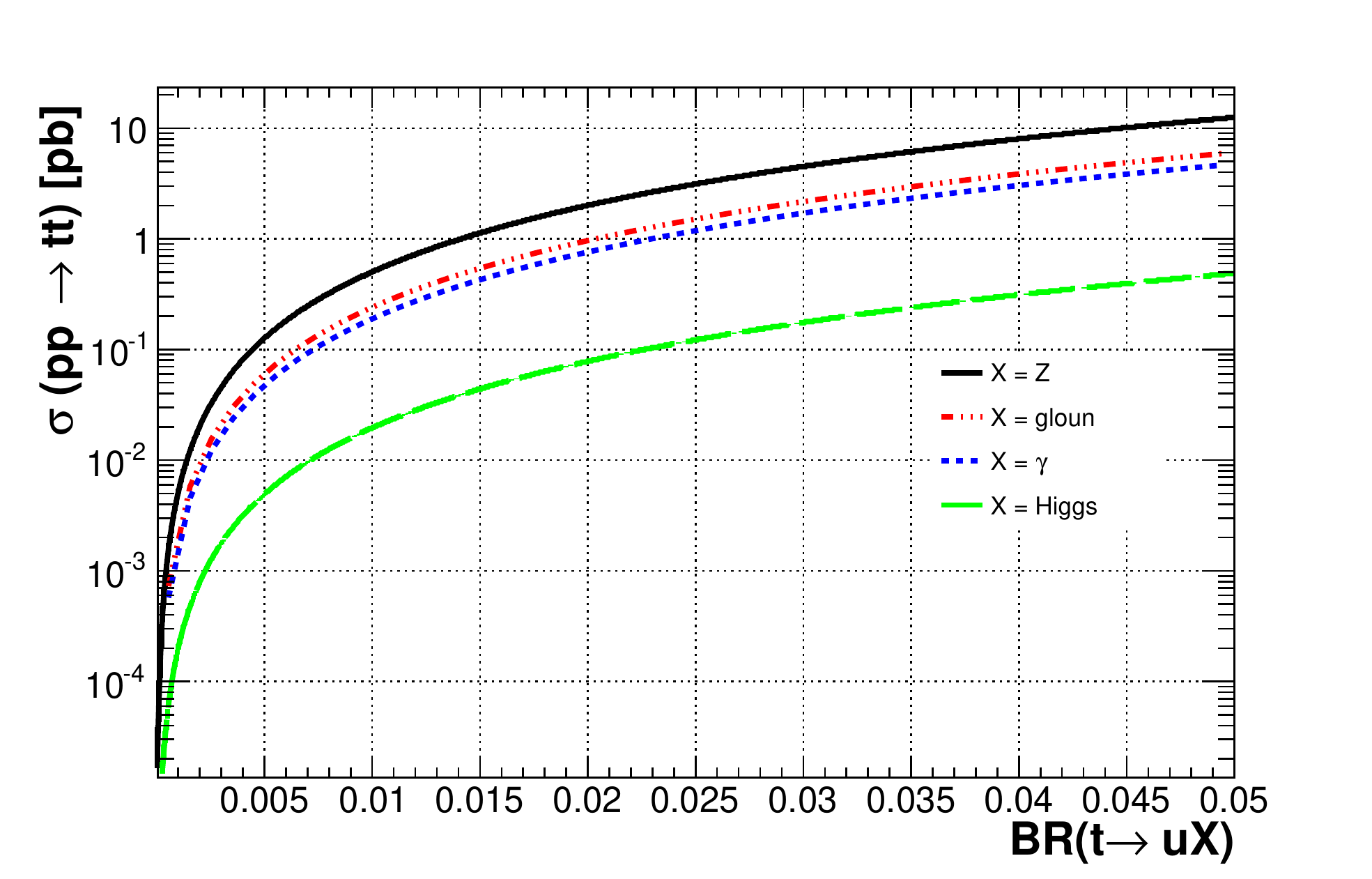} \\
\hfill

\caption{\label{fig:i} Anomalous top pair production cross section for the process $pp \rightarrow tt$ due to $tuX$ anomalous vertices versus the FCNC branching ratios for the decays BR$(t\rightarrow u\gamma)$, BR$(t\rightarrow uZ)$, BR$(t\rightarrow ug)$ and BR$(t\rightarrow uH)$.    }
\label{sigmavsbr}
\end{figure}

In Fig. \ref{sigmavsbr} we plot the cross section of the anomalous production of same-sign top quarks for the LHC at $8$ TeV against the branching ratio of the top-quark FCNC decays. We show the $tt$ cross section originating from different anomalous interactions separately. The red curve corresponds to the $tug$ anomalous coupling while other anomalous couplings are set to zero. As can be seen, any bound on the production of  same-sign top quarks (which is available from the LHC results) immediately implies a bound on the anomalous top FCNC decays. Another interesting observation from Fig. \ref{sigmavsbr} is the relative sensitivity of the FCNC top-quark decays due to a photon, $Z$-boson, gluon, or Higgs exchange to same-sign top pair production.

Four separated samples of 100 000 events are generated independently corresponding to anomalous $tt$ production through FCNC interactions. In the production of the signal events, the top-quark branching ratio to a bottom quark and a $W$ boson is assumed to be 100\%. Then the $W$ boson is required to decay only into a charge lepton ($e$, $\mu$, or $\tau$) and a neutrino in {\sc MadGraph} to ensure good statistical coverage and include leptonic tau decays. {\sc pythia} \cite{pythia} is used to simulate the subsequent showering and hadronization effects. Detector effects are simulated using {\sc delphes} \cite{delphes}. The {\sc delphes} card for simulating the CMS detector is modified in order to include the lepton and $b$-tag efficiencies calculated by the CMS Collaboration as discussed in the previous section.

We analyze each signal channel separately. The CMS same-sign lepton search is closely followed to determine the efficiency for FCNC signal events passing the selections. Similar cuts are applied on the selected leptons, jets, $b$ jets, $H_T$, and $E_T^{\text{miss}}$ from simulated signal samples.

\section{Results}
\label{sec:4}

The same-sign dilepton final state coming from $t \rightarrow Wb \rightarrow l \nu b$ in same-sign top production is associated with two $b$ jets and missing transverse energy  from the undetected neutrinos. In addition, the signal samples are dominated by the events with positive charged leptons, as discussed in the previous section. Therefore, the most sensitive signal region  
in our search for FCNC interactions is SR2. In this category, the signal efficiency is high, while the SM backgrounds and their uncertainties are small compared to other signal regions. In other words, the best significance is obtained from SR2.

As no excess above the SM expectation is observed, the $95\%$ C.L. upper bound on the number of new-physics events were  set in Ref. \cite{cmstt}. In Table \ref{expinput} the bounds are shown in each of the nine signal regions. In order to determine more conservative upper bounds, the results considering a $30\%$ uncertainty on the signal efficiency are chosen between $10, 20$ and $30\%$.  

The results for the signal region SR2 are used to set the limit on the FCNC anomalous couplings. In the derivation of the limits, the contributions of $tq\gamma$, $tqZ$, $tqg$, and $tqH$ to same-sign top production are considered separately. Therefore, the limits are evaluated on one of the FCNC couplings, while setting the other couplings to zero. The limits on the strength of FCNC anomalous couplings can be converted to limits on the anomalous top-quark decays and are summarized in Table \ref{myresult}.

\begin{table*}[ptbh]
\centering
\begin{tabular}{|c|c|c|c|c|}
\hline
Process ($tqX$) & $\kappa_{uX} (\kappa_{cX}=0)$ & $\kappa_{cX} (\kappa_{uX}=0)$&  ${\cal B}R(t\rightarrow uX) (\%)$  &  ${\cal B}R(t\rightarrow cX) (\%)$  \\
\hline 
$tq\gamma$ & $0.23$ &  $1.28$&  $1.27$ &  $38.15$ \\
\hline
$tqZ$ & $0.20$ &  $1.15$&  $0.80$ &  $25.52$ \\
\hline
$tqg$ & $0.05$ &  $0.25$&  $1.02$ &  $27.73 $ \\
\hline
$tqH$ & $0.39$ &  $1.30$&  $4.21$ &  $45.46$ \\
\hline
\end{tabular}
\caption{\label{tab:i} The observed 95\% upper limits on the top-quark FCNC anomalous
couplings and branching ratios.
}
\label{myresult}
\end{table*}

\begin{figure*}[tbp]
\centering 
\includegraphics[height=0.30\linewidth,angle=00]{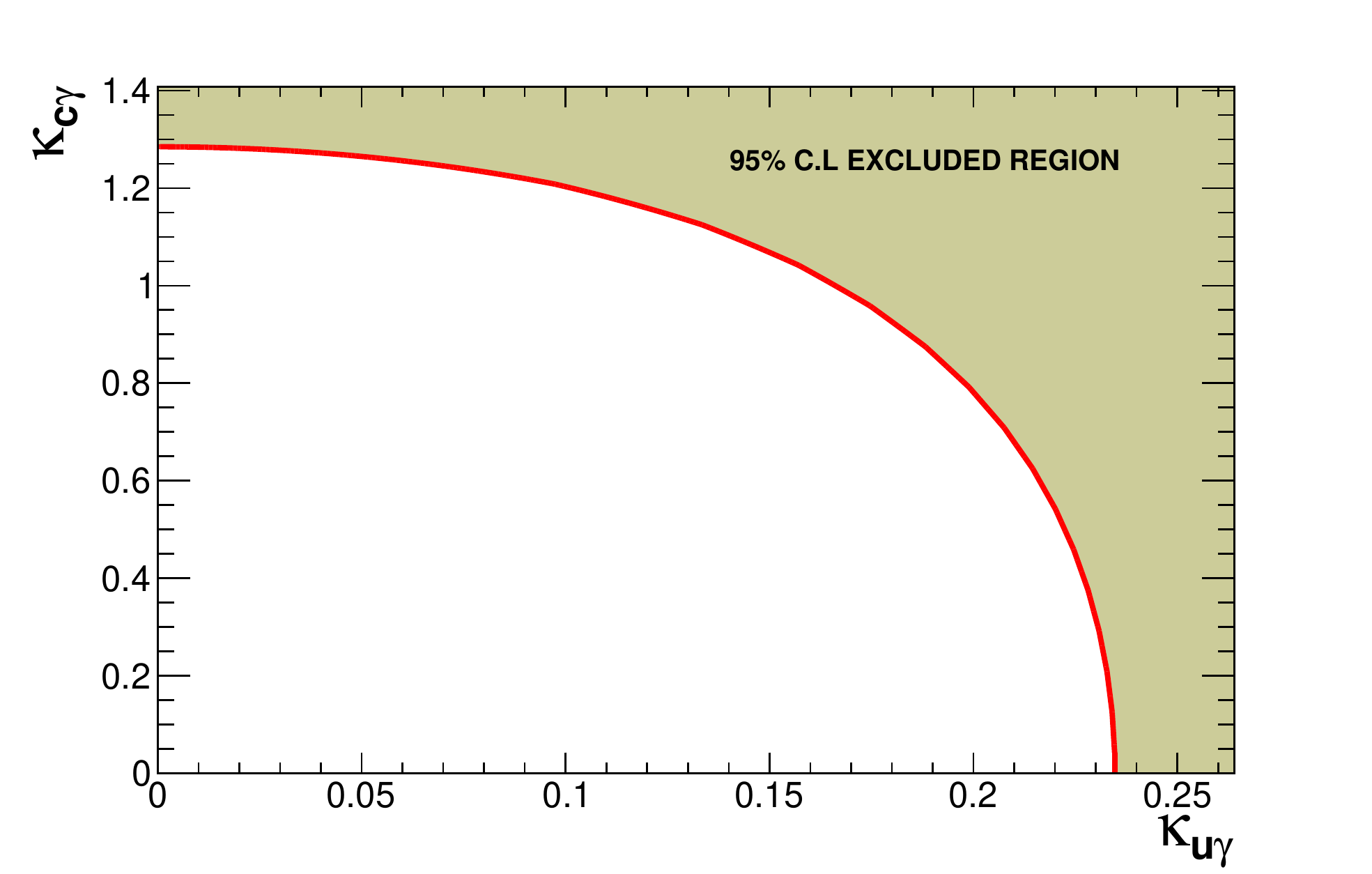}
\hfill
\includegraphics[height=0.30\linewidth,angle=00]{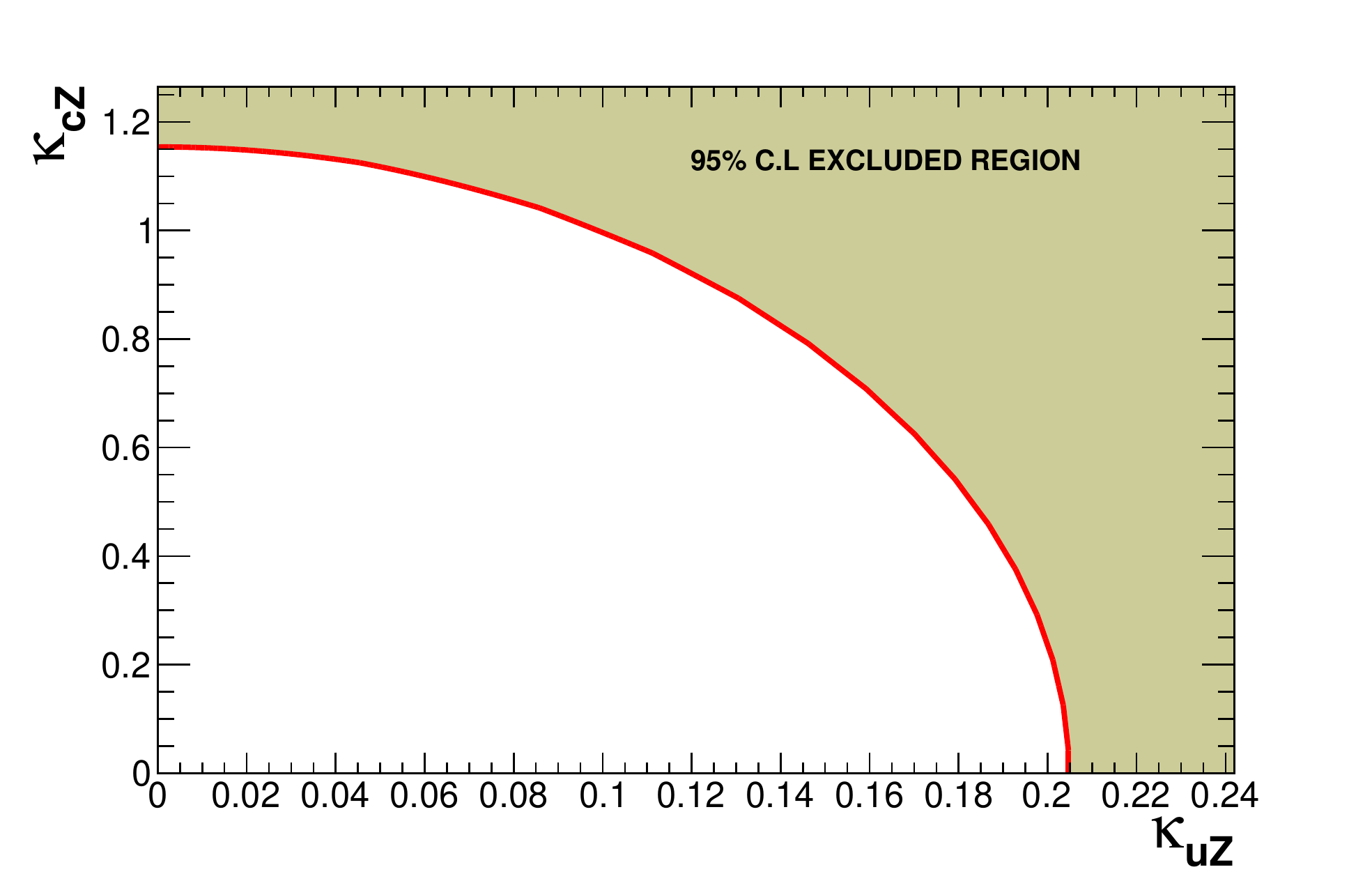}
\hfill
\includegraphics[height=0.30\linewidth,angle=00]{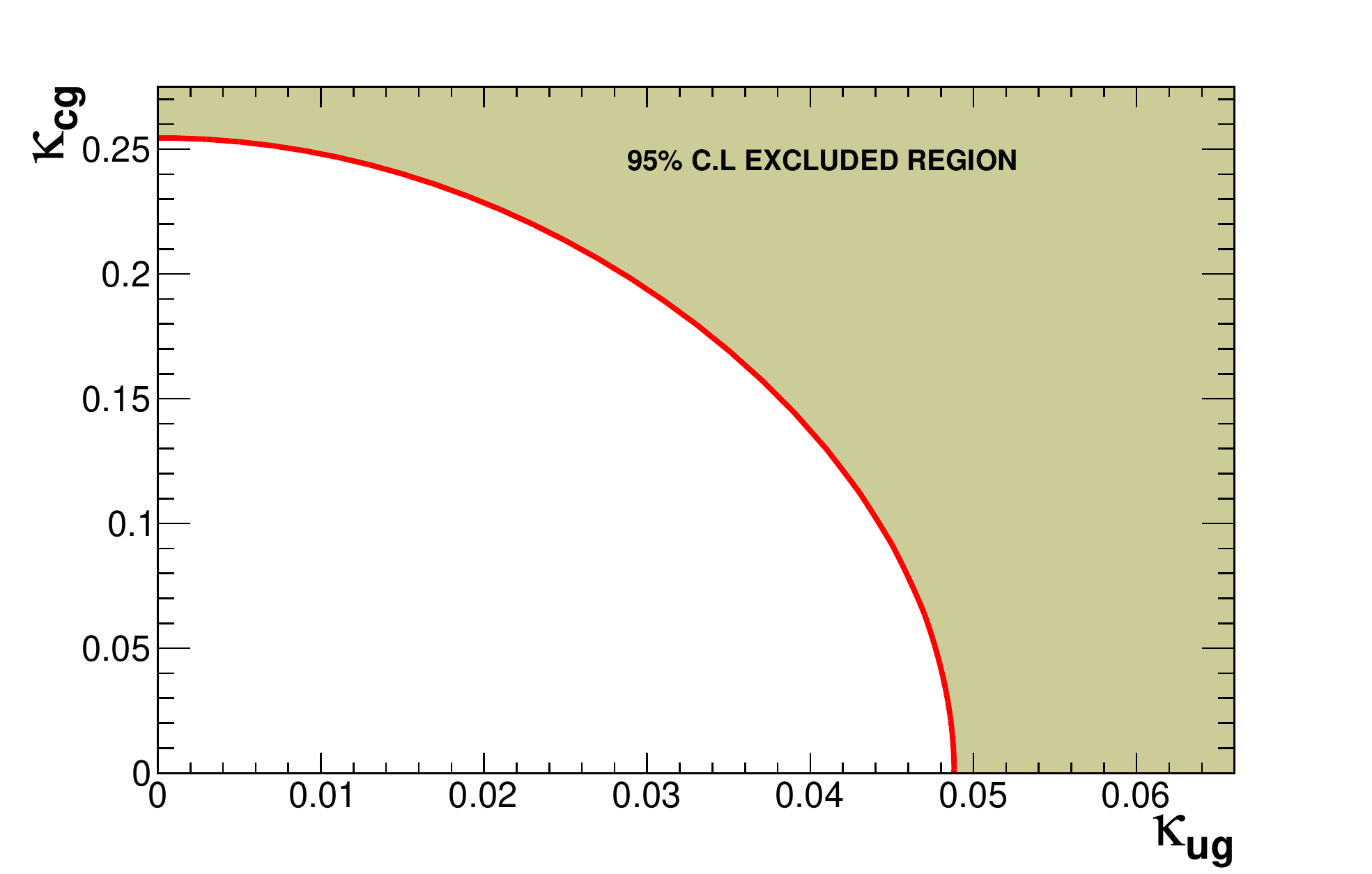}
\hfill
\includegraphics[height=0.30\linewidth,angle=00]{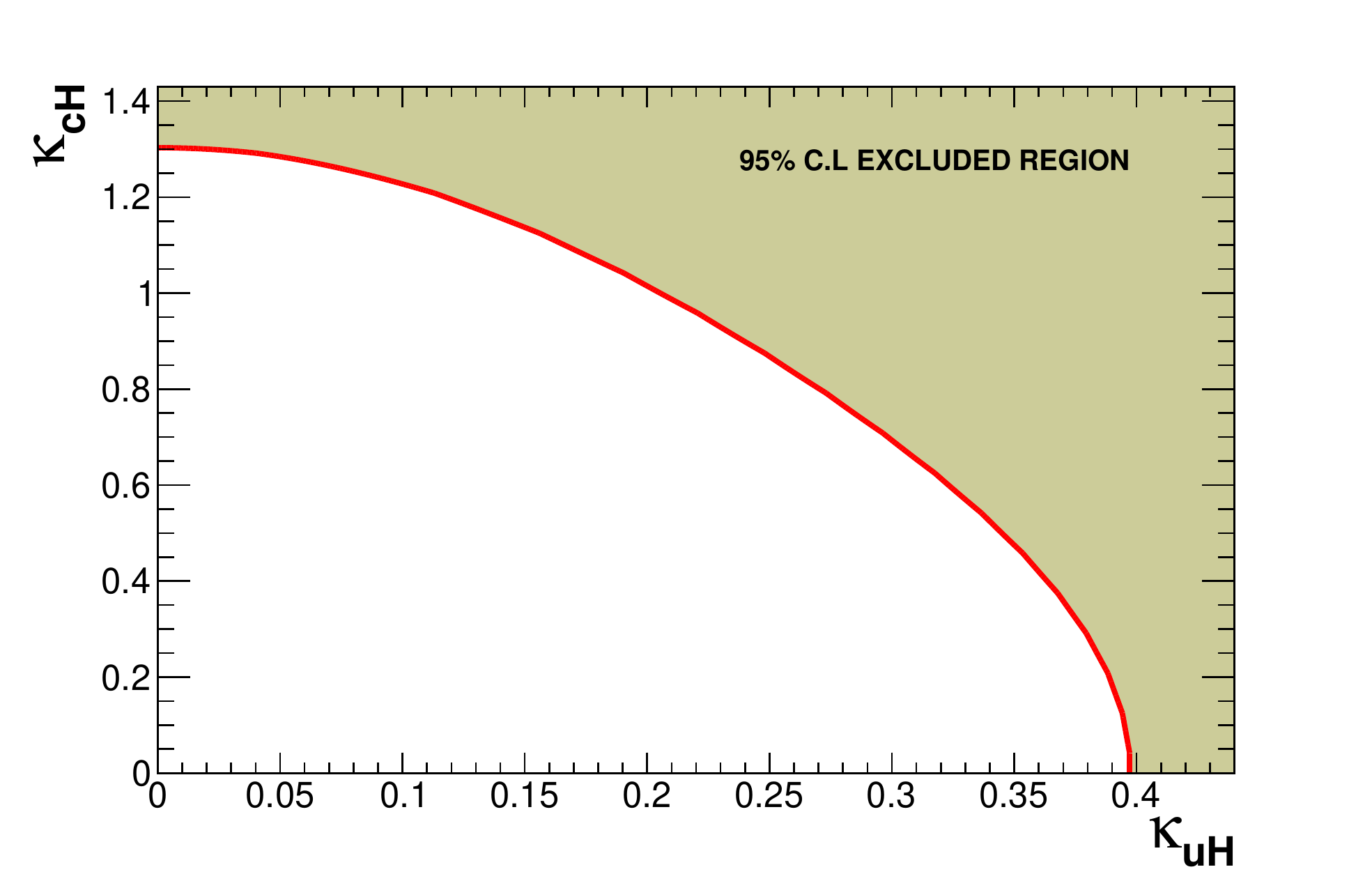}
\caption{\label{fig:i} Excluded region at 95\% C.L. in the $\kappa_{uX}-\kappa_{cX}$ plane for $X= \gamma$ , $Z$, gluon, and Higgs.}
\label{kappalimit}
\end{figure*}

\begin{figure*}[tbp]
\centering 
\includegraphics[height=0.30\linewidth,angle=00]{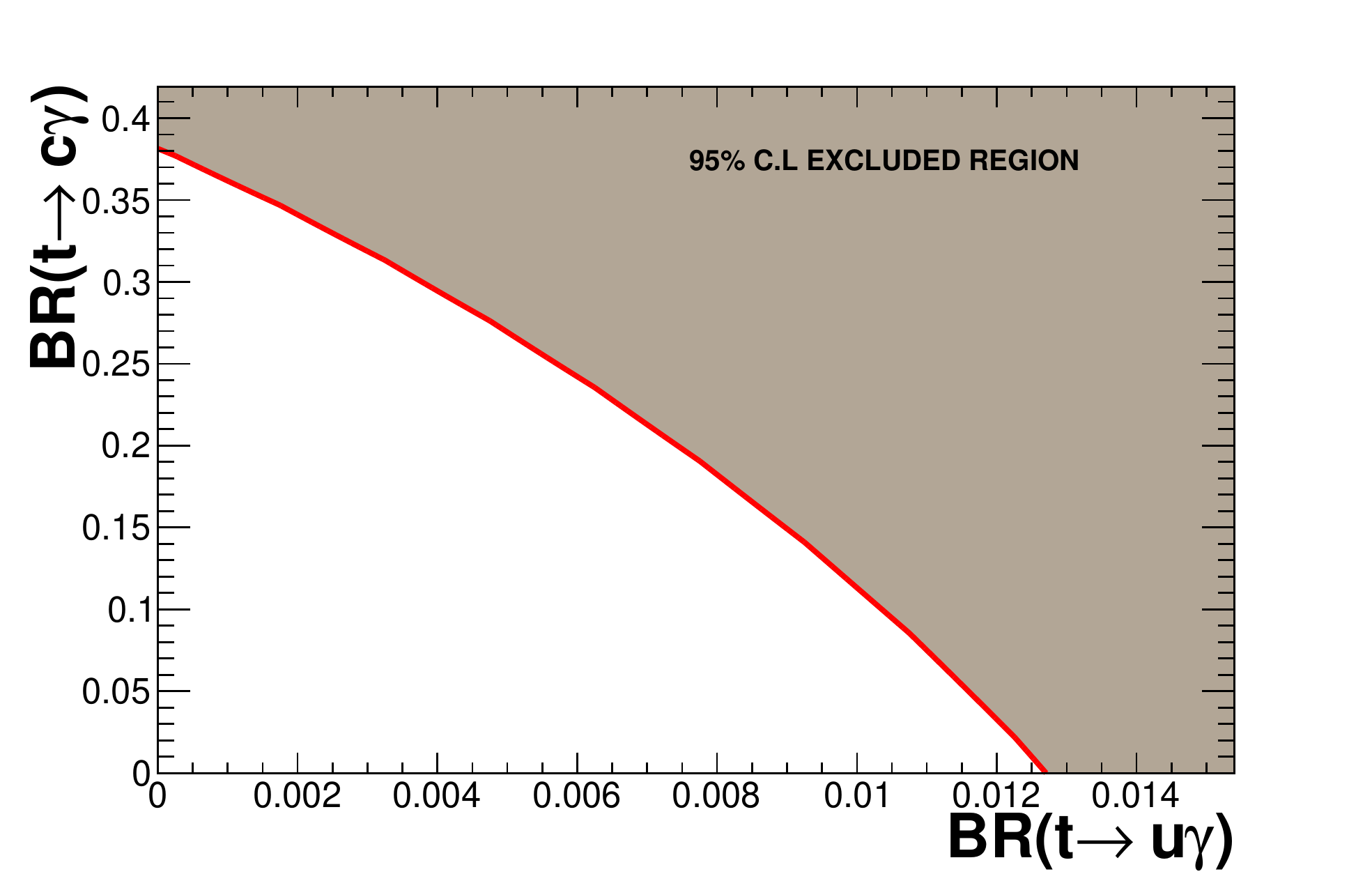}
\hfill
\includegraphics[height=0.30\linewidth,angle=00]{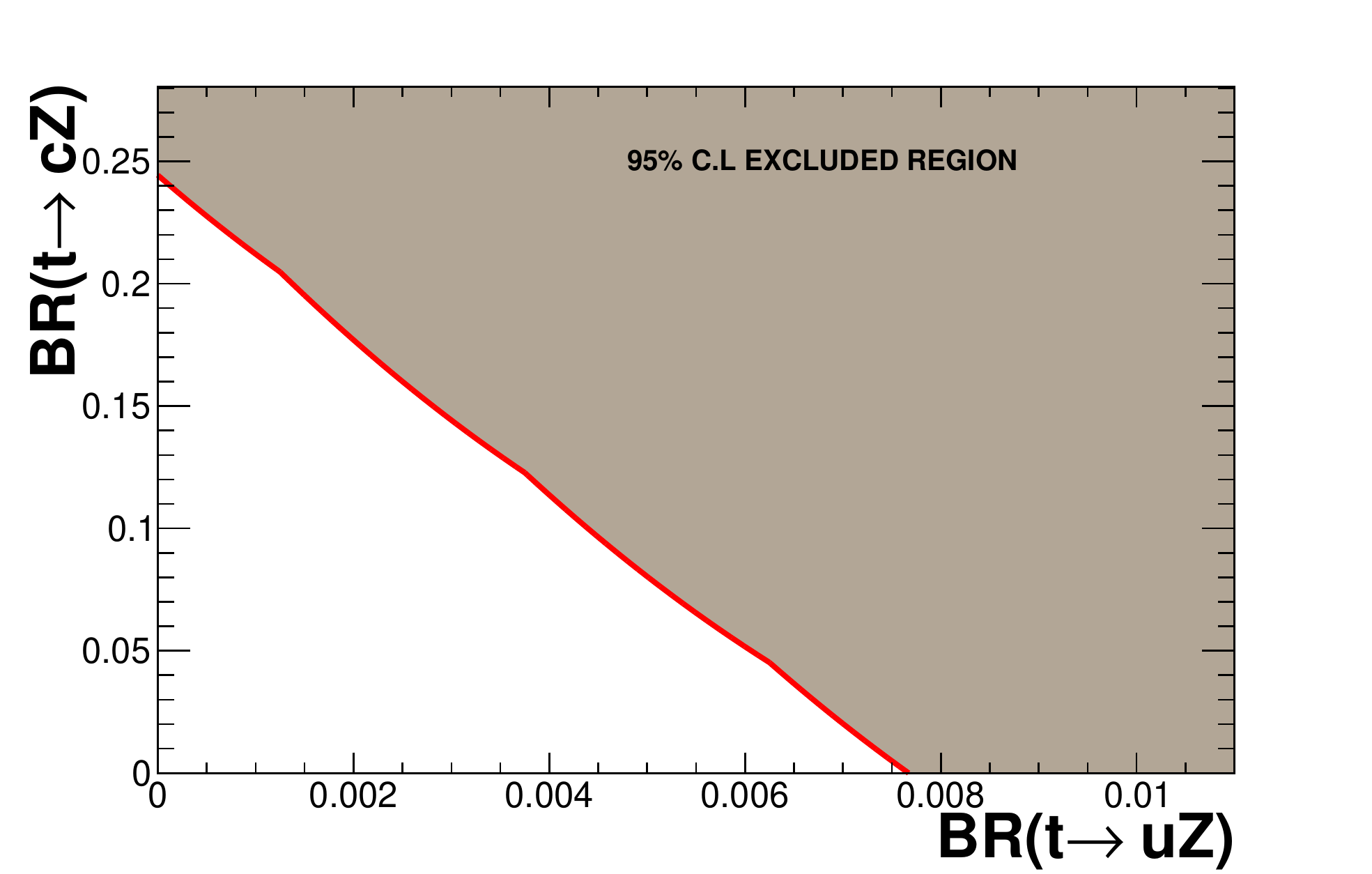}
\hfill
\includegraphics[height=0.30\linewidth,angle=00]{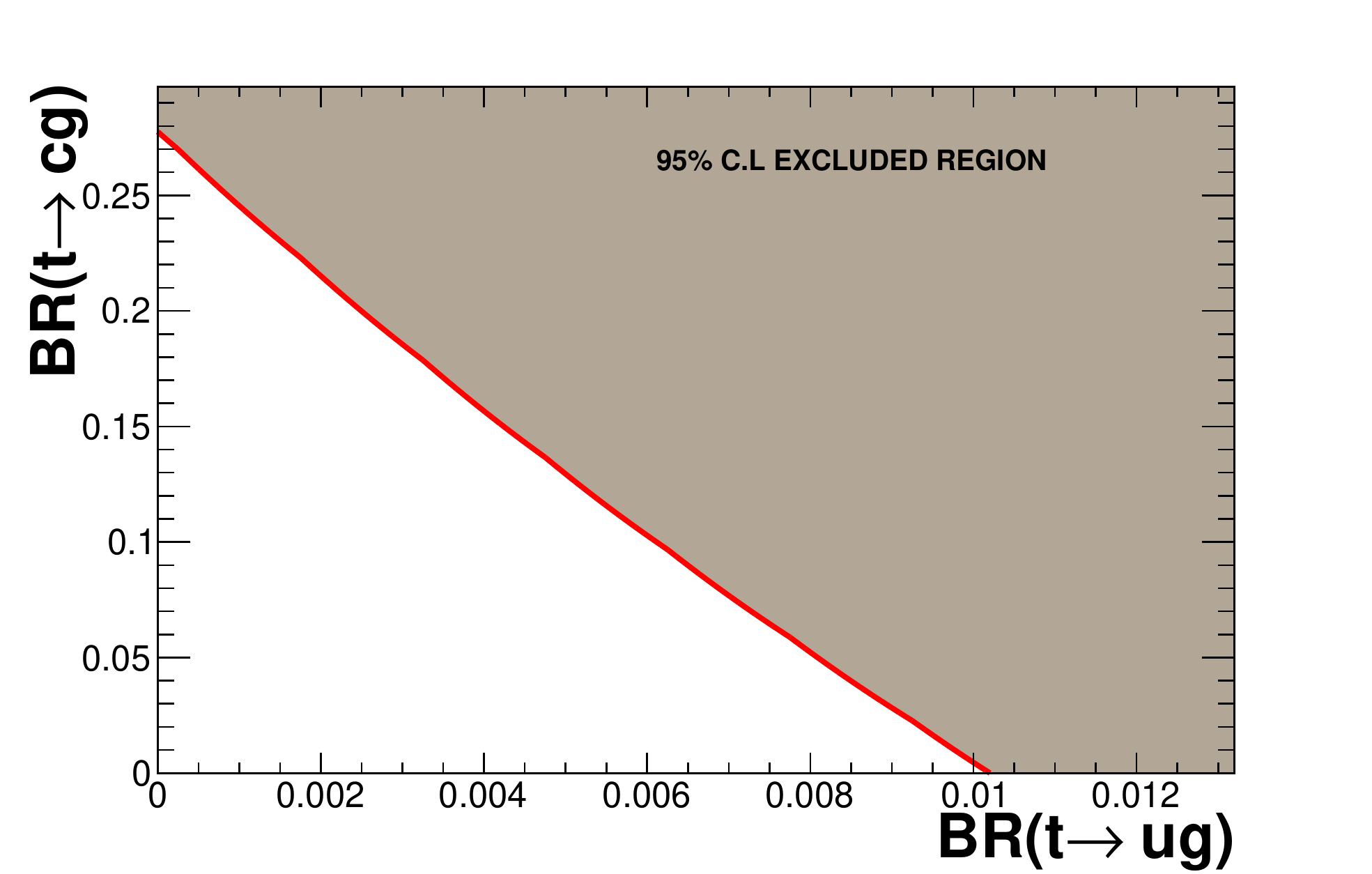}
\hfill
\includegraphics[height=0.30\linewidth,angle=00]{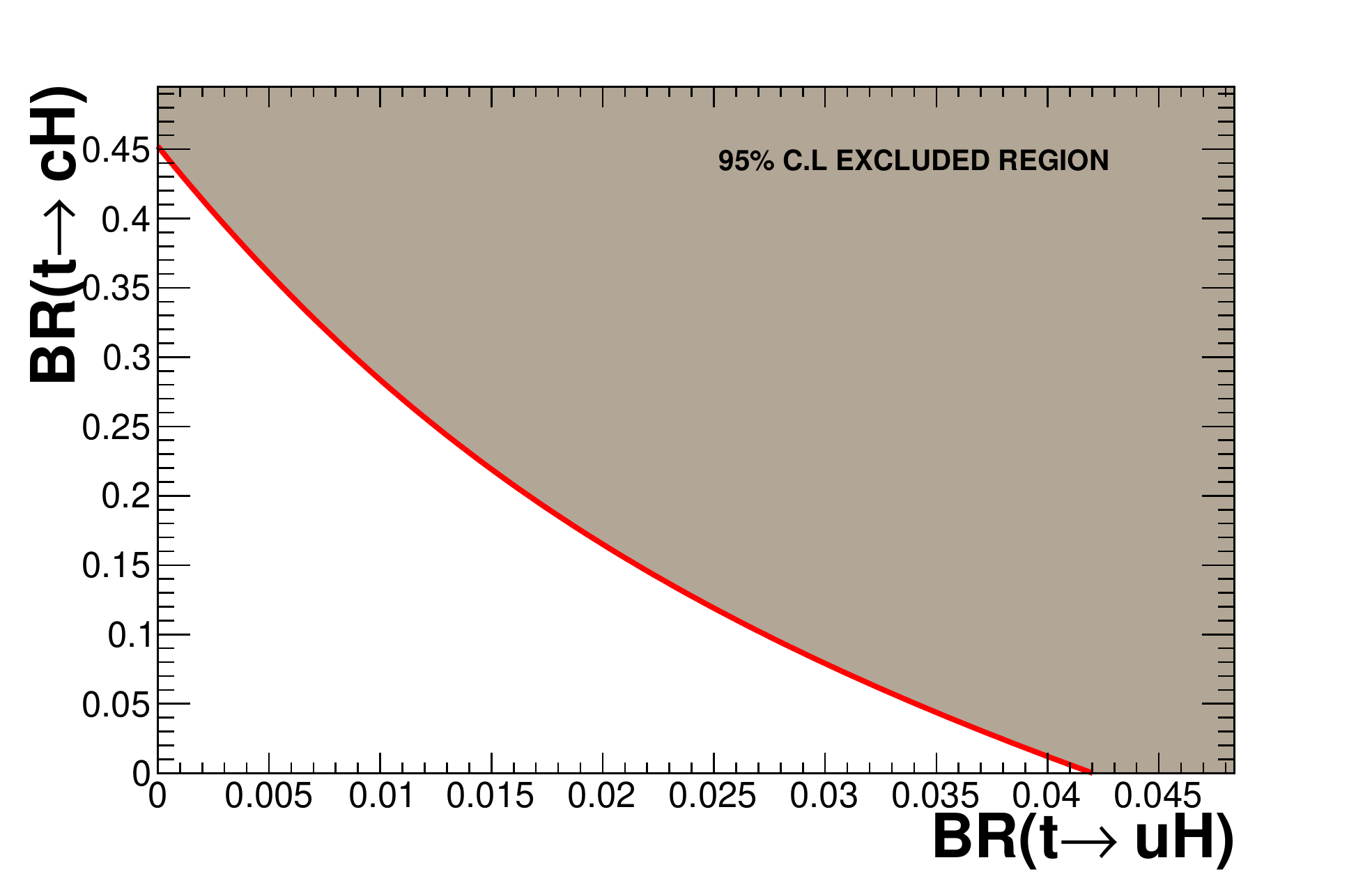}
\caption{\label{fig:i} Excluded region at 95\% C.L. in the BR$(t\rightarrow uX)- \text{BR}(t\rightarrow cX)$  plane for $X= \gamma$ , $Z$, gluon, and Higgs.}
\label{brlimit}
\end{figure*}

Figures \ref{kappalimit} and \ref{brlimit} show  the 95\% C.L. excluded region in the ($\kappa_{uX},\kappa_{cX}$) and (BR$(t\rightarrow uX),\text{BR}(t\rightarrow cX)$) plane obtained by this analysis. Due to the PDF of the proton, the LHC data are less sensitive to  the $\kappa_{cX}$ parameter than  $\kappa_{uX}$.

\section{Sensitivity at the 14 TeV LHC}
\label{sec:5}

\begin{figure}[tbp]
\centering 
\includegraphics[width=0.6\textwidth]{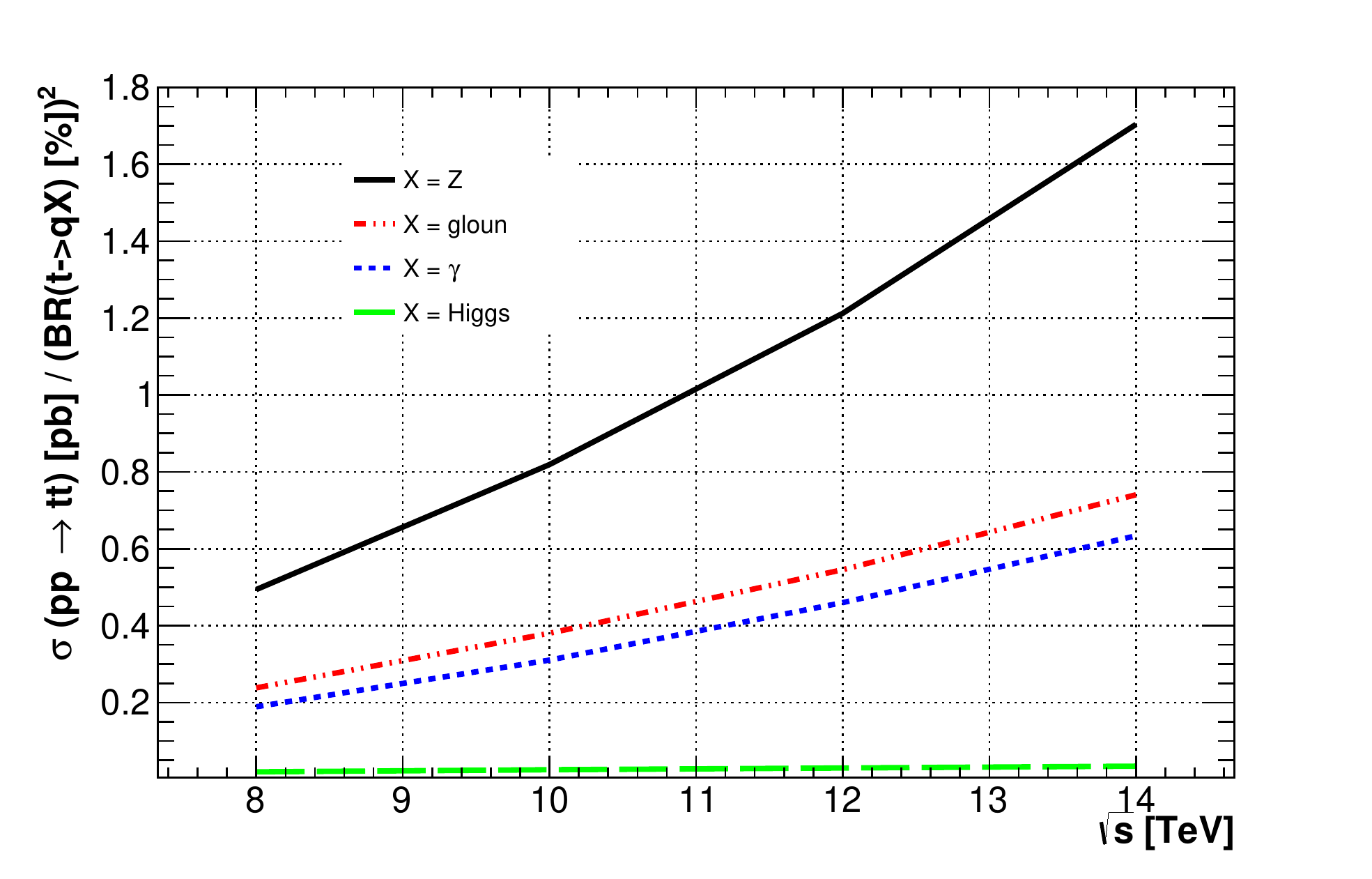} \\
\caption{\label{fig:i} Anomalous top-pair production cross section for the process $pp \rightarrow tt$ due to $tuX$ anomalous vertices divided by the square of the FCNC branching ratios for the decays BR$(t\rightarrow q\gamma)$, BR$(t\rightarrow qZ)$, BR$(t\rightarrow qg)$ and BR$(t\rightarrow qH)$ versus the center-of-mass energy.    }
\label{com}
\end{figure}

In this section, we study the sensitivity of future searches for FCNC interactions through same-sign top-quark production. In Fig. \ref{com}, the cross sections for $tt$ production induced by flavour-violating top-$Z$, top-photon, top-gluon, and top-Higgs couplings normalized to the corresponding top-quark decay branching ratios are shown as a function of the center-of-mass energy. We find from Fig. \ref{com} that for a given value of $BR(t\rightarrow qX)$ FCNC branching ratios, the anomalous $tZq$ coupling is the most sensitive coupling, followed by $tqg$, $tq\gamma$, and $tqH$ in the shown range of the center-of-mass energy. The cross sections of all signal channels increase by increasing the center of mass energy and the cross section due to the fact that tqH increases less than the others.

The same-sign analysis used to constrain the top-quark FCNC interactions has three sources of SM background:
\begin{itemize}
\item Fake leptons, i.e., the selected leptons do not originate from either from the decay of a boson, or the decay of a $\tau$ lepton. 
\item Charge flips:,i.e., the electron charge is mismeasured due to severe bremsstrahlung in the tracker material.
\item Rare SM processes which mostly come from $t\bar{t}W$ and $t\bar{t}Z$.
\end{itemize}

The contributions of these three sources are reported separately in Ref.  \cite{samesignDilepton}. The simulation does not properly reproduce the contribution of the backgrounds with fake or charge-flipped leptons (instrumental background). Therefore, data-driven methods were used to estimate the instrumental background contribution from data in Ref. \cite{samesignDilepton}. In addition to the considerable contribution of these backgrounds, they are also the main source of uncertainties. However, this makes it impossible to precisely predict the expected limit for higher center-of-mass energies or an arbitrary luminosity. 

To estimate the reach of the search for FCNC anomalous production of same-sign top quarks at the 14 TeV LHC using the results of a search identical to the CMS search at 8 TeV \cite{samesignDilepton}, we need to estimate the contribution of the instrumental and rare SM backgrounds at the 14 TeV LHC.
The method (which was developed in Sec. 3 of Ref \cite{susysamesign})is used to estimate the contribution of instrumental backgrounds. The idea is as follows. Due to the final selection criteria, (specially two the $b$-jets requirement), $t\bar{t}$ is the main source of instrumental background. So one can scale the rate of these backgrounds with the $t\bar{t}$ cross section approximately and predict their contributions at 14 TeV. 
To estimate the contribution of the rare SM backgrounds, we first produce the $t\bar{t}W$ and $t\bar{t}Z$ samples at 14 TeV using {\sc MadGraph5} \cite{madgraph}. We then use {\sc pythia} \cite{pythia}and {\sc delphes} \cite{delphes} to simulate the showering, hadronization, and detector effects with the same condition explained in Sec \ref{sec:32}. Finally the same cuts as mentioned before are imposed to find the number of events from $t\bar{t}W$ and $t\bar{t}Z$ in SR2. 

The prediction of the expected backgrounds at 14 TeV is validated by calculating the ratio of the instrumental background to the rare SM backgrounds for SR6 and SR8 and comparing them with the ratios calculated in Ref \cite{susysamesign}. The ratios are fully compatible in both regions, and it is calculated to be 2.02 for SR2.

\begin{figure*}[tbp]
\centering 
\includegraphics[height=0.30\linewidth,angle=00]{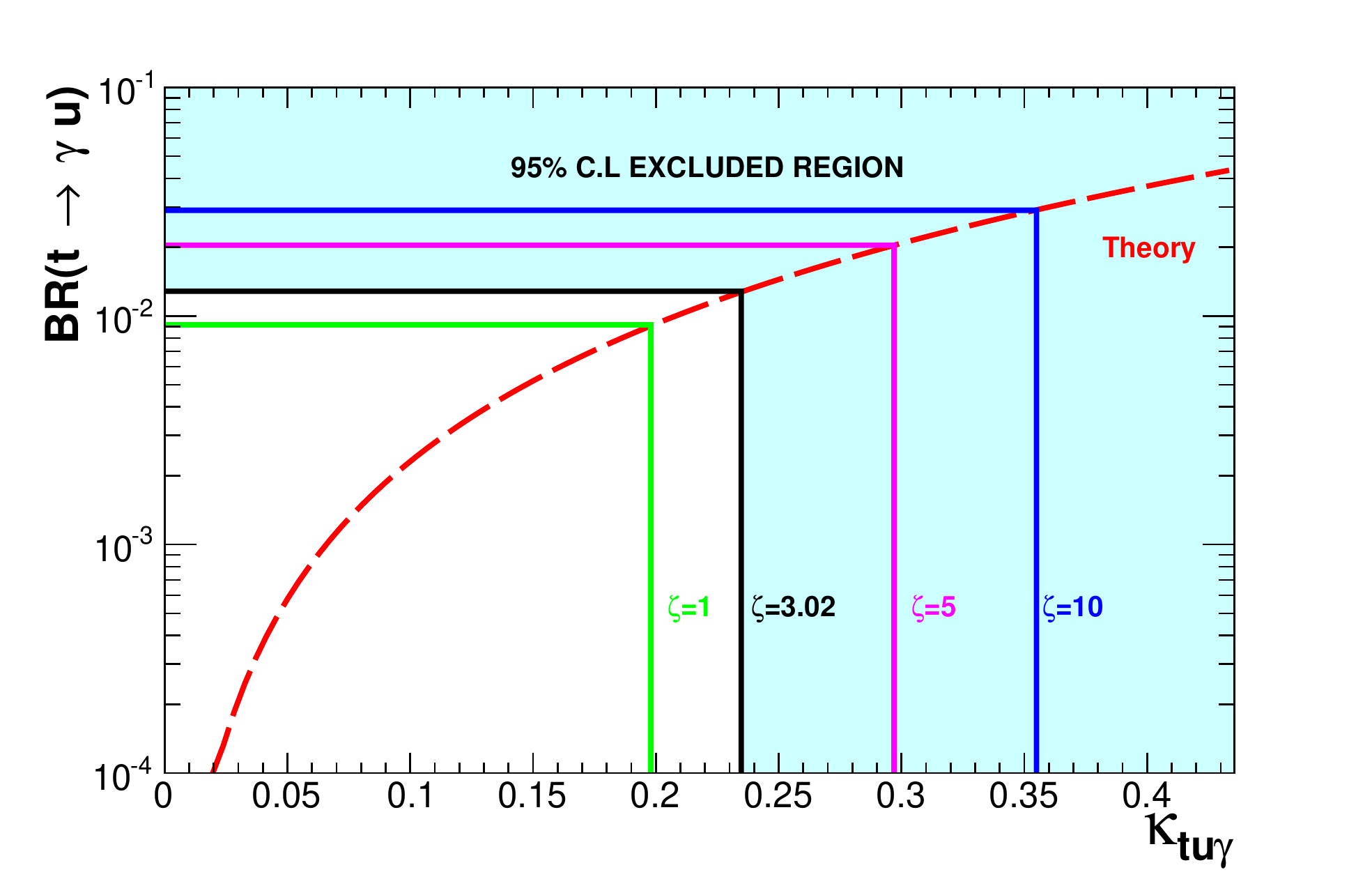}
\hfill
\includegraphics[height=0.30\linewidth,angle=00]{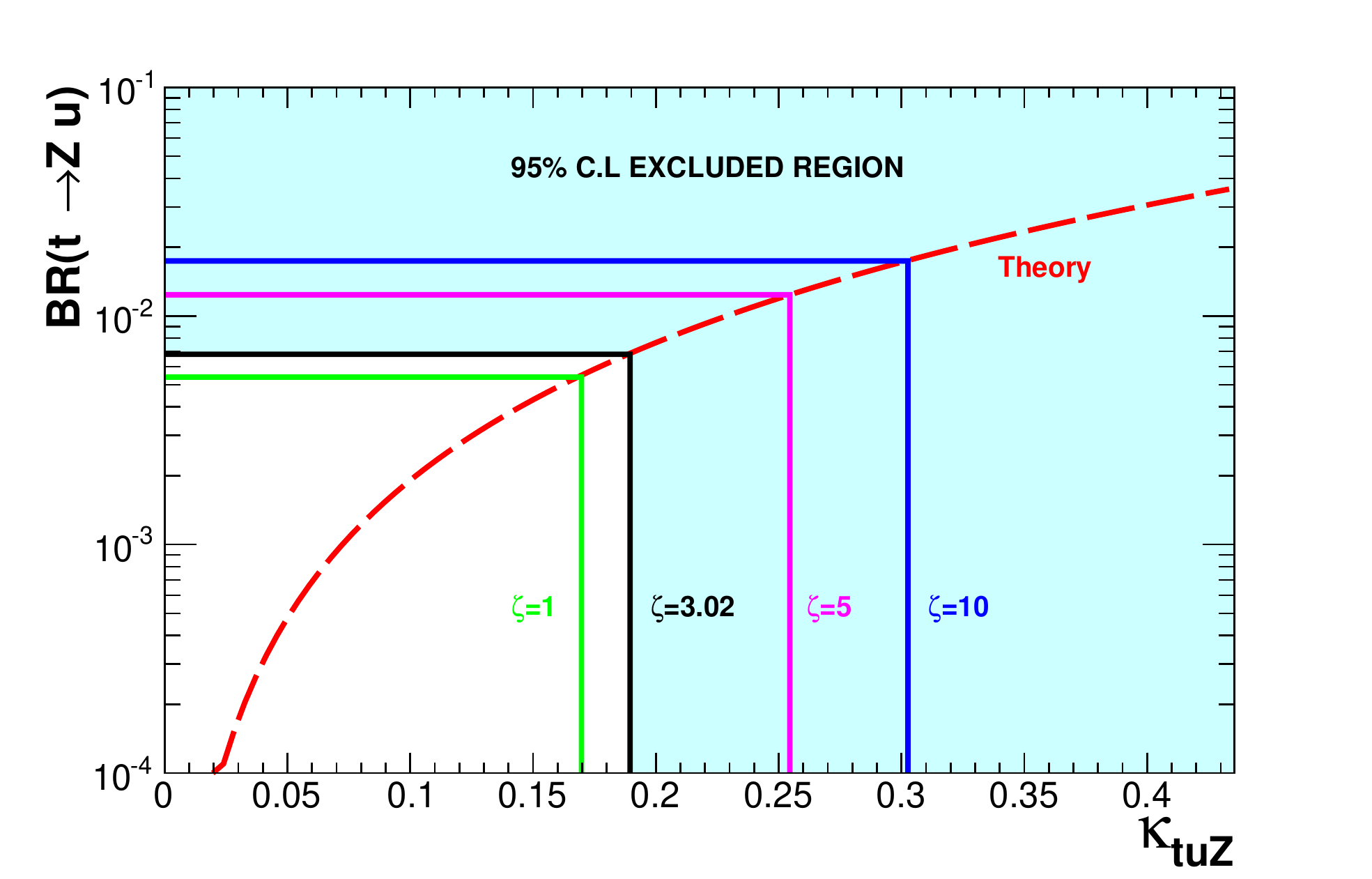}
\hfill
\includegraphics[height=0.30\linewidth,angle=00]{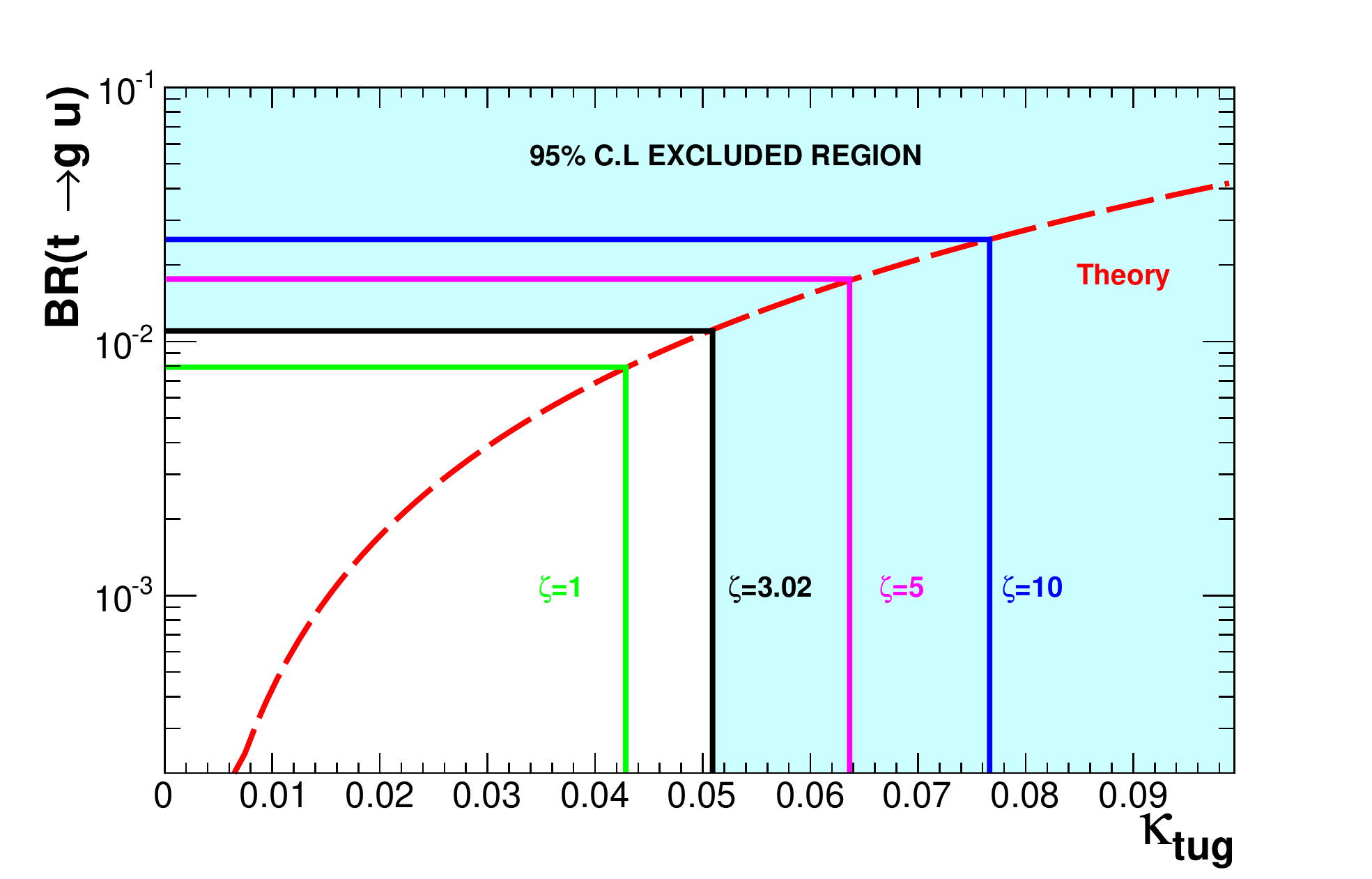}
\hfill
\includegraphics[height=0.30\linewidth,angle=00]{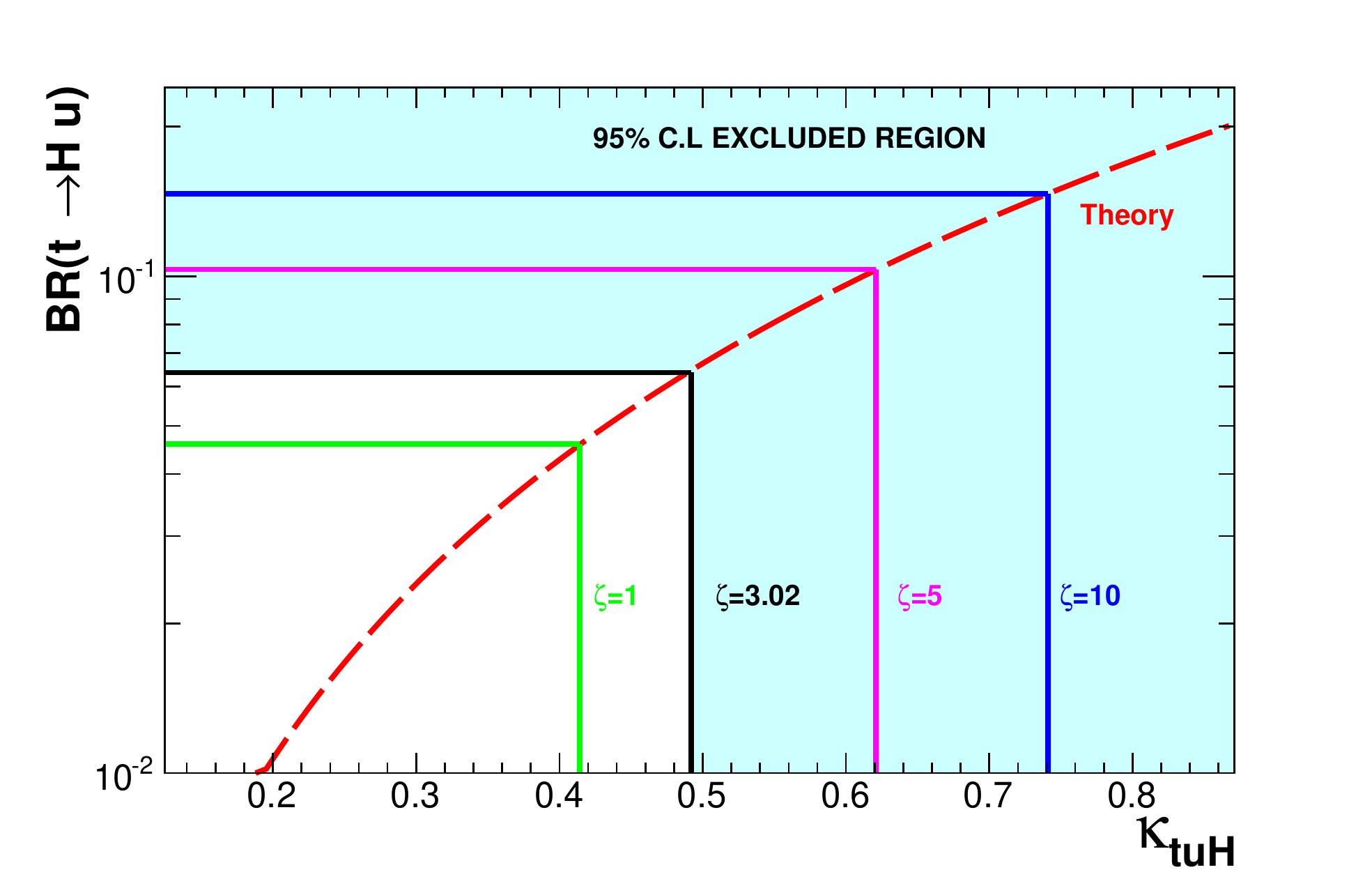}
\caption{\label{fig:i} 
Estimated 95$\%$ CL expected exclusion reach of flavour violating top-Z, top-photon, top-gluon and top-Higgs couplings and related top quark branching ratio through the same-sign top channel signature at the 14 TeV LHC with 100 $fb^{-1}$. Concerning the important effect of the instrumental background on the predicted results, a range is assumed for $\zeta$. $\zeta=3.02$ is obtained by rescaling the experimental result in SR2 at 8 TeV. The theoretical prediction for the top quark FCNC branching ratios versus the anomalous couplings are also shown. }
\label{PRElimit}
\end{figure*}

In order to estimate the uncertainties of instrumental backgrounds, no detector simulation is performed. The uncertainty obtained from 8 TeV is scaled according to the $t\bar{t}$ cross section to evaluate the uncertainty at 14 TeV, which leads to a large uncertainty. It would make our analysis more useful if we could present the results considering different uncertainties from instrumental backgrounds. In order to illustrate the uncertainty effects, we define
\begin{eqnarray}
\label{zeta}
\zeta &=& \frac{{\rm Total \ background \ rate}}{{\rm Irreducible \ background \ rate}}\nonumber\\ &=& 1 + \frac{{\rm Instrumental \ background \ rate}}{{\rm Irreducible \ background \ rate}}
\end{eqnarray}

where both rates are calculated after all cuts. Combining the rare SM and instrumental backgrounds, the 95\% predicted exclusion limits are presented in Fig. \ref{PRElimit} for all signal channels. The $\zeta$ is varied between 1 to 10 to show how the reach of the analysis would change by changing the uncertainty on the instrumental backgrounds. Using the nominal value for $\zeta=3.02$, it can be seen that the 95\% excluded region boundaries have not improved significantly at the 14 TeV LHC compared to the 8 TeV LHC for the $tu\gamma$, $tuZ$, and $tug$ signal channels, and the reach is even worse for $tuH$. 

The CMS Collaboration has updated the same-sign analysis with 19.5 fb$^{-1}$ of data \cite{samesignupdate}. The total uncertainties are increased in different signal regions by increasing the luminosity from  10.5 fb$^{-1}$ to 19.5 fb$^{-1}$. Our studies show that using updated experimental results would not change the results obtained in this analysis.   This behavior confirms that scaling the instrumental background rates and their related uncertainties to the $t\bar{t}$ cross section is a good approximation.

\section{Discussion and Conclusions}
\label{sec:6}

In this work, we have analyzed the same-sign top-quark pair signature of the
top-quark flavour-changing neutral interactions through photon, $Z$ boson, gluon, and Higgs boson exchanges in proton-proton collisions. The experimental results obtained by CMS at a center-of-mass energy of 8 TeV were used to constrain the top-quark anomalous couplings and branching ratios.

Whereas the limits obtained on the the FCNC branching ratios of top-quark decays are found to be noncompetitive compared to recent results derived from anomalous single top-quark production or the anomalous decay of a top quark in $t\bar{t}$ events \cite{cmsgamma,cmsz,atlasg,cmsh,atlash,cmsh2,tchmultilepton,thuth}, these results provide an interesting cross-check of the evidences for the absence of the top-quark FCNC interactions in a different physics process. 

The limits could be improved if the CMS Collaboration updates the same-sign dilepton search by subdividing the signal regions exclusively, so the results from different signal regions could be combined. 
An improvement of the lepton and $b$-tag efficiencies and the systematic uncertainties on the background predictions would improve the results.

Using the results of the CMS same-sign dilepton search at 8 TeV \cite{samesignDilepton}, we tried to predict the possible reach of the 14 TeV LHC. However, the presence of the instrumental backgrounds as an important source of uncertainties makes the prediction of the analysis reach vague. We find that with selections identical to the 8 TeV search no significant improvement is expected to be obtained for the top-quark FCNC process through the same-sign dilepton signature at the 14 TeV LHC. 

\section*{Acknowledgments}
The author thanks Mojtaba Mohammadi Najafabadi for useful discussions and comments on the manuscript.


\end{document}